# Simulating Organogenesis in COMSOL: Phase-Field Based Simulations of Embryonic Lung Branching Morphogenesis

Lucas D. Wittwer[1], Roberto Croce[1,2], Sebastian Aland[3,4] and Dagmar Iber[*1,2]
[1]D-BSSE, ETH Zurich, Switzerland, [2]Swiss Institute of Bioinformatics (SIB), Switzerland,
[3]Institute of Scientific Computing, TU Dresden, Germany
[4]Faculty of Informatics/Mathematics, HTW Dresden, Germany
*Corresponding author: D-BSSE, ETH Zurich, Mattenstrasse 26, 4058, Switzerland,
dagmar.iber@bsse.ethz.ch

**Abstract:** Organogenesis has been studied for decades, but fundamental questions regarding the control of growth and shape remain unsolved. We have recently shown that of all proposed mathematical models only ligand-receptor based Turing models successfully reproduce the experimentally determined growth fields of the embryonic lung and thus provide a mechanism for growth control during embryonic lung development. Turing models are based on at least two coupled non-linear reaction-diffusion equations. In case of the lung model, at least two distinct layers (mesenchyme and epithelium) need to be considered that express different components (ligand and receptor, respectively). The Arbitrary Lagrangian-Eulerian (ALE) method has previously been used to solve this Turing system on growing and deforming (branching) domains, where outgrowth occurs proportional to the strength of ligand-receptor signalling. However, the ALE method requires mesh deformations that eventually limit its use. Therefore, we incorporate the phase field method to simulate 3D embryonic lung branching with COMSOL.

## 1. Introduction

The development of an organism from a single cell, the fertilized oocyte, involves innumerous symmetry breaks that need to occur at the right time and the right place to give rise to a functional organism. During mouse lung development, thousands of branches form in a highly stereotyped process [1, 2]. The underlying mechanism that repetitively guides such a highly deterministic patterning and growth process has long fascinated biologists and theoreticians [3].

In 1952, Alan Turing described a reaction-diffusion mechanism that can give rise to such deterministic symmetry breaks [4]. 20 years later Gierer and Meinhardt [5] as well as Prigogine [6] independently defined chemical reactions that are consistent with such a Turing patterning mechanism. Turing mechanisms have since been suggested for a wide range of biological patterning phenomena [7]. While experiments have confirmed the Turing mechanism in chemical reaction systems [8, 9], their proof in a biological system is still outstanding. In fact, alternative mechanisms are more likely in several systems where Turing mechanisms have previously been proposed [10, 11].

The increasing availability of quantitative imaging data offers new opportunities to test and challenge proposed Turing mechanisms [12]. We have previously shown that only a ligand-receptor based Turing mechanism [13], but none of the other proposed alternative mechanisms, can reproduce the measured embryonic growth fields during early lung development [14]. In the lung, the ligand-receptor based Turing mechanism is based on the interaction of the ligand ($L$) FGF10 with its receptor ($R$) FGFRIIb [14, 15] as FGF10 signalling has been shown to be both necessary and sufficient for branch formation [16-20]. Accordingly, we compared the measured embryonic growth fields to the predicted concentration fields of the ligand-receptor complex ($R^2L$).

The ligand and receptor dynamics can be described by the following set of non-dimensionalized partial differential equations (PDEs) where the ligand $L$ is defined in the domain $\overline{\Omega}_{\text{mes}}$ surrounding the epithelium (red in Fig. 1), while the receptor $R$ is defined only on the cell surface of the epithelial layer, $\Gamma_\Omega$ [13, 15, 21-23] (Fig 1).

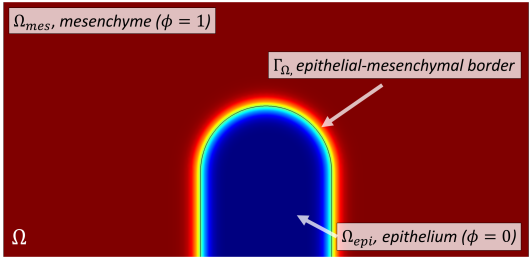

**Figure 1: 2D Setting**. The mesenchymal bulk $\Omega_{mes}$ (red) encloses the epithelium (blue). The ligand $L$ is defined in $\overline{\Omega}_{mes}$ only. The receptor $R$ is defined on the epithelial-mesenchymal border $\Gamma_\Omega$.

$$\frac{\partial R}{\partial t} = \Delta R + \gamma(a - R + R^2 L) \quad \text{on } \Gamma_\Omega \quad (1)$$
$$\frac{\partial L}{\partial t} = D\,\Delta L + \gamma b \quad \text{in } \Omega_{mes} \quad (2)$$
$$D\vec{n} \cdot \nabla L = -\gamma R^2 L \quad \text{on } \Gamma_\Omega \quad (3)$$

Here, in Eqs. 1 and 2, the terms on the left hand side denote the time derivatives, and the first term on the right hand side are the diffusion terms. $D > 1$ refers to the relative diffusion coefficient of ligand and receptor, and $\gamma$ is a scaling constant that defines the relative speed of the reaction terms. $a$ and $b$ refer to the constitutive receptor and ligand production rates. The epithelium expresses the receptor, while the mesenchyme produces the ligand. The ligand can diffuse throughout the domain, while the receptor can diffuse only in the epithelial layer. Receptors are internalized constitutively at rate $-R$, while ligand is removed mainly upon receptor binding at rate $-R^2 L$ (boundary condition described in Eq. 3). A key requirement of the Turing mechanism is that ligand binding triggers receptor accumulation on the cell surface at rate $R^2 L$. FGF10-dependent receptor recycling to the cell surface is indeed observed [24].

Further analysis of the Turing mechanism in branching morphogenesis requires the simulation of the 3D branching process over time. This is numerically challenging because the growth and deformation of two distinct tissue layers, the branching epithelium and the surrounding mesenchyme, needs to be coupled to a changing concentration field.

## 2. Phase Field Method

Many numerical techniques are available to handle the coupling of surface and bulk equations on evolving geometries [25]. In particular, the Arbitrary Lagrangian-Eulerian (ALE) method has been used to solve this Turing system on growing and deforming (branching) domains, but meshing problems develop over time that limit its use [14, 15, 22, 26-28]. For the problem considered here, interface-capturing methods are advantageous since they can handle arbitrarily complex geometries without re-meshing. In these methods, the surface is implicitly described as a level-set of an auxiliary field variable $\phi$. The most popular methods of this kind are the level-set method [29-31] and the phase field method, which we choose here.

The phase field method has a long history in the theory of phase transitions dating back to van der Waals [32]. The phase field $\phi$ describes the physical phases by taking distinct values in each of the domains (e.g. $\phi = 0$ within the epithelium and $\phi = 1$ outside the epithelium) with a smooth transition in between, around the interface (Fig. 2). Hence, the interface is diffuse with a finite width $\epsilon$, and an intermediate level set of the phase field (e.g. $\phi = 0.5$) may be used to get a discrete interface location.

A phase field method to simulate partial differential equations (PDEs) in complex evolving geometries has been presented in [33] for bulk geometries and in [34] for surfaces. In a biological context, the approach has been used to solve equations on complex geometries such as trabecular bone [35] and to describe vesicle budding during endocytosis [36]. The coupling of surface and bulk equations has been presented in [37]. The method is capable to model transport, diffusion, reaction and adsorption/desorption of arbitrary material quantities on deformable surfaces. The approach has been used to model two-phase flows with soluble nanoparticles [38] and soluble surfactants [39].

The general idea to use a phase field for coupled bulk and surface equations is intriguingly simple: the integrals of the weak form of the equations are multiplied by characteristic functions, which allows to extend the integration domains to a larger computational box $\Omega$ (see Fig. 1). For example, multiplication by $\phi$ can be used to extend integrals defined on the epithelium exterior to a surrounding box. Analogously, the approximate Dirac delta function $|\nabla \phi|$ can be used for integration along the epithelium surface. As a result, the phase field enters the corresponding equations. Since the receptor $R$ is defined only on the epithelial boundary, we multiply the integrands in the weak form of Eq. 1 with $\delta = |\nabla \phi|$ which leads (in strong from) to

$$\delta \frac{\partial R}{\partial t} = \nabla \cdot (\delta \nabla R) + \gamma \delta (a - R + R^2 L) \quad \text{in } \Omega.$$

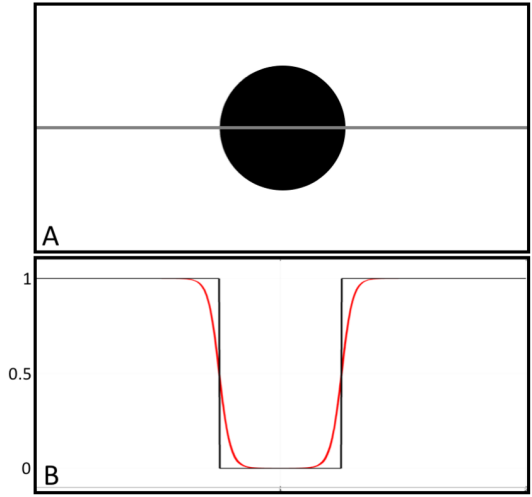

**Figure 2: Difference between a sharp and a diffuse interface representation. (A)** A black circle embedded in a rectangle containing a cut-line. **(B)** The profile along the cut line (grey) in panel A is shown for a sharp interface as would be used in an ALE approach (black), and for a smooth transition between $\phi = 0$ and $\phi = 1$, corresponding to the phase field transition (red).

Matched asymptotic analysis shows convergence towards Eq. 1 as $\epsilon$ tends to zero [34]. To increase numerical stability, we introduced some further modifications. First, to ensure that the interaction term is restrained to the interface we multiply the $R^2 L$ term with

$$\chi(\phi) = \begin{cases} 1 & 0.01 < \phi < 0.99 \\ 0 & elsewhere. \end{cases}$$

Second, $R$ is constantly extended off the epithelial interface in normal direction similar to [36] by including the term $D_n \nabla \cdot (\delta \vec{n} \vec{n} \cdot \nabla R)$, where $\vec{n} = \nabla \phi / |\nabla \phi|$ represents the outward normal on the epithelial boundary. In summary, Eq. 1 becomes

$$\delta \frac{\partial R}{\partial t} = \nabla \cdot (\delta \nabla R) + \gamma \delta (a - R + \chi(\phi) R^2 L) + D_n \nabla \cdot (\delta \vec{n} \vec{n} \cdot \nabla R) \quad \text{in } \Omega. \quad (4)$$

Similarly, we restrict Eq. (2) to the mesenchyme by multiplication with $\phi$. To include the boundary condition given in Eq. 3, we limit the interaction term, $R^2 L$, with the $\delta$-function to the epithelial boundary, as discussed for Eq. 1. We do not need the stabilizing $\chi$-function in Eq. 2 because the non-linear term dampens the evolution of $L$. Thus, the diffuse-domain version of Eq. 2 reads

$$\phi \frac{\partial L}{\partial t} = D \nabla \cdot (\phi \nabla L) + \gamma (\phi b - \delta R^2 L) \quad \text{in } \Omega. \quad (5)$$

In this way Eq. 5 approaches the sharp interface equation for interface thickness $\epsilon \to 0$ [33].

Advection of the phase field can be realized by a stabilized level-set equation [40]. In our model, lung growth depends on the $R^2 L$-concentration and is directed in the normal direction of the epithelium boundary. Therefore, we evolve the phase field with the concentration dependent velocity field

$$\vec{v}_{growth} = s \cdot R^2 L \cdot \vec{n}, \quad (6)$$

where $s$ is a scaling factor.

## 3. Use of COMSOL Multiphysics® Software

We use COMSOL Multiphysics® to perform all simulations and post-processing tasks. The initial lung geometries were obtained from embryonic lung images and were imported as STL-files through the CAD Import Module. In the previous ALE-simulations we employed the Coefficient Form PDE-, the Surface Reaction- and the Moving Mesh-module to simulate the growing embryonic lung with Eqs. 1-3, 6. A more detailed description can be found in [27, 28, 41].

To circumvent the problems that arise in the ALE-implementation due to the highly displaced mesh, we implement the modified Eqs. 4 and 5 instead of Eqs. 1-3. COMSOL offers two different types of phase fields, one based on the Cahn-Hilliard equation and one based on a level-set approach, described in [40]. The Cahn-Hilliard implementation is available in the Phase Field module. It defines the two phases by $\phi = -1$ and $\phi = 1$ respectively, allowing values slightly below $-1$ and above $1$. The level-set approach is available in the Level-Set module and defines the phases by $\phi = 0$ and $\phi = 1$. For this study we used the Level-Set module as the governing equation exhibits less unwanted self-dynamics, even though the Cahn-Hilliard equations can be solved more efficiently (data not shown). We define the phase field as $\phi = 0$ in the epithelium and $\phi = 1$ within the mesenchyme, such that the continuous interface in-between represents the epithelial-mesenchymal border (see Fig. 1, 2). To overcome over- and undershoots and to smooth $\phi$ at the two ends of the interface we further transformed $\phi$ as

$$\varphi(\vec{x}) = \max(\min(\phi(\vec{x}), 1 - 10^{-5}), 10^{-5}),$$

and used this modified $\varphi$ in Eqs. 4, 5. Similarly, we have to set a lower bound for $\delta$,

$$\tilde{\delta}(\vec{x}) = \max(\delta(\vec{x}), 10^{-5}).$$

The last modification affects the resulting velocity field $\vec{v}_{growth}$. By multiplying $\vec{v}_{growth}$ with $\tilde{\delta}$ we focus the displacement on the geometry interface at $\phi = 0.5$. The reaction-diffusion process on the epithelium as well as the diffusion in the mesenchyme bulk are modelled with the Coefficient Form PDE-module.

## 4. Results

We first solved the system of equations for $L$, $R$ and $\phi$ on a highly regular 2D geometry (extended Fig. 1) with the Level-Set module. With $D = 100, a = 0.15, b = 0.1, \gamma = 0.1$ and $D_n = 100$ on a quadratic domain of size 200, we obtain the plausible stable patterning shown in Fig. 3A, without requiring the $\chi(\phi)$ stabilisation function. The geometry was meshed with the COMSOL standard settings for "Extremely fine". The re-initialization parameter $\gamma$ of the Level-Set-module was set to $\gamma = 0.1$ and the interface thickness was half the maximum mesh element size in the region of the interface ($\epsilon_{ls} = $ ls. hmax/2). Enabling growth after reaching this stable pattern (with the scaling factor $s = 0.001$) leads to several branching events, resulting in the deformation shown in Fig. 3B-D. The growing geometry fills the domain over time without touching the boundaries. This behaviour arises because the ligand $L$ is produced only in the space between the branching epithelium and the bounding box, and its concentration becomes too low to support outgrowth as the epithelial boundary (that acts as a sink for ligand) approaches the bounding box. In Fig. 3D on the right, two branches grow together. This is not a physiological behaviour of the embryonic lung, but the fusion of branches has been observed in the developing pancreas.

Figure 4 shows the concentration fields of $R$ and $L$, which are defined in the whole domain (Fig. 4A,B), as well as their reaction domains (Fig. 4C,D) at time $t = 15000$. The effect of the stabilization term for $R$, which extends the concentration constantly in normal direction, can be seen in Fig. 4A. The restriction of $R$ to the epithelial boundary via multiplication with $\delta$ is

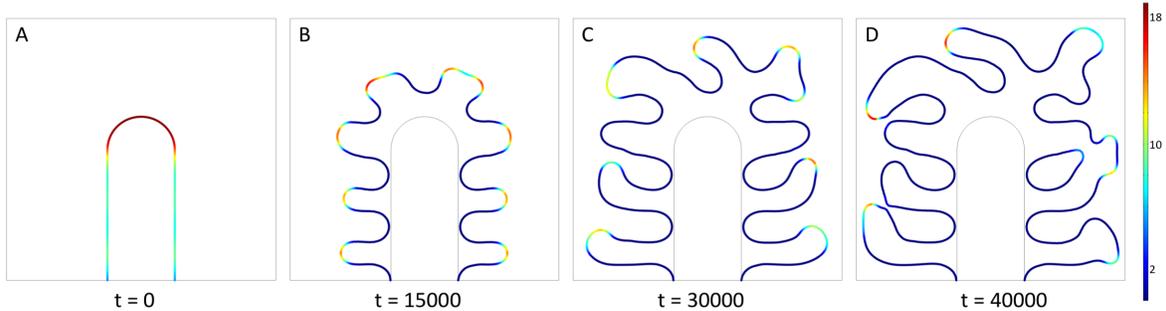

t = 0    t = 15000    t = 30000    t = 40000

**Figure 3: 2D branching behavior**. The phase field is indicated by the $\phi = 0.5$ contour-line and colored based on the value of $R^2 L$ (colour bar). **(A)** Stationary pattern on the initial geometry. **(B-D)** Enabling growth leads to deforming outgrowth. (B) Until $t = 15000$, growth is axis-symmetric, but (C) irregularities occur afterwards. (D) The space-filling behavior is based on the decreasing $L$ concentration at the boundaries of the domain. The merging of two branches is not observed in the development of the lung.

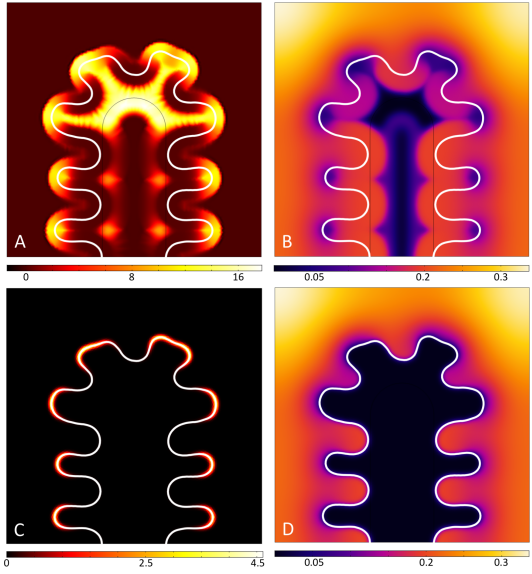

**Figure 4: Concentration of R and L at t = 15000. (A, B)** Concentration of (A) $R$ and (B) $L$, which both exist in the whole domain. **(C, D)** The concentration domains of active $R$ and $L$. (C) $R$ is restricted to the interface via multiplication with $\delta$, while (D) $L$ is restricted to its reaction domain via multiplication with $\phi$. The white contour-lines are at $\phi = 0.5$.

shown in Fig. 4C. Similarly, the actual reaction domain of $L$, that is obtained by multiplying the concentration of $L$ with $\phi$, is shown in Fig. 4D.

Finally, we used the phase field method with a 3D embryonic lung geometry (Fig. 5). Contrary to the 2D case we need the $\chi(\phi)$ stabilisation function. On a static domain we again obtain stable patterns (Fig. 5A); the parameters for the Turing mechanism are the same as in 2D, except for $\gamma = 0.2$. The mesenchyme geometry is meshed with the COMSOL standard settings of "Extra fine". The level set re-initialization parameter is $\gamma = 0.3$ and the same interface thickness based on the mesh size as in 2D was chosen. Growth is, however limited by the small size of the mesenchyme, the bounding box (grey part in Fig. 5A), which does not grow in our simulations. To demonstrate the potential of the method to handle strong deformations of the epithelium, we repeat the simulations with an artificially 1.5-fold enlarged mesenchyme (compare grey parts in Fig. 5A,B). In this case, the spatial pattern (Fig. 5B) changes because the $L$-secreting region is now bigger and thus the concentration of $L$ changes. Enabling growth with $s = 0.015$ leads to more pronounced spots and prolonged branch outgrowth (Fig. 5C,D). As in the 2D simulations, some branches fuse (black arrow in Fig. 5D).

## 5. Conclusions

The phase-field module in COMSOL offers a robust method to solve models where domain growth and deformation is coupled to dynamic reaction-diffusion reactions, whose concentration domains are restricted to dynamic geometric subdomains. We have implemented and tested the method with a Turing model for lung branching morphogenesis, and we obtain similar patterns as with the ALE-approach. As the complexity of the geometry increases over time, the phase field approach, however, seems not to suffer from numerical instabilities. We will use the phase field method in our future research to understand how size and shape are controlled during development, and during branching morphogenesis in particular.

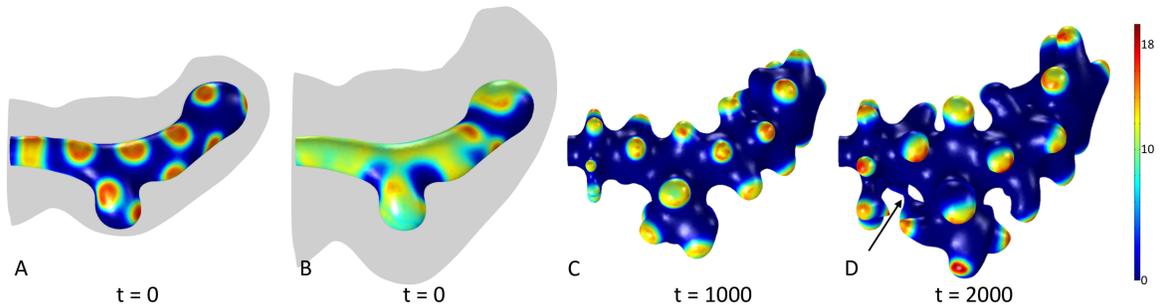

**Figure 5: Turing patterns on 3D embryonic lung geometries and concentration-based growth. (A)** Turing pattern on the epithelial surface, represented by the $\phi = 0.5$ isosurface. The receptor $R$ is defined on this surface only. The ligand $L$ is defined in the grey area. **(B-D)** Turing pattern and branch outgrowth when the mesenchyme is enlarged 1.5-fold. The resulting patterns differ due to a bigger domain where the ligand $L$ is produced. In (D), two branches grow together (indicated by the black arrow). This would not occur during embryonic lung development.

# 7. Acknowledgements

We thank our colleagues for discussion and the COMSOL support staff for their excellent support, especially Sven Friedel and Zoran Vidakovic. SA acknowledges support from the German Research Foundation through grant no. AL-1705/1.